\documentclass[twocolumn,nofootinbib,prd,aps,superscriptaddress,tightenlines]{revtex4}
\usepackage{graphicx}

\def\nn{\nonumber \\ }
\def\abs#1{\left| #1 \right|}

\begin{document}

\preprint{ \vbox{\hbox{CALT-68-2585} 
  \hbox{UCSD/PTH 06-02}}}

\title{Modifications to the Properties of the Higgs Boson} 

\author{Aneesh V. Manohar}
\affiliation{Department of Physics, University of California at San Diego,
9500 Gilman Drive, La Jolla, CA 92093-0319}

\author{Mark B. Wise}
\affiliation{California Institute of Technology, Pasadena, CA 91125}

\date{January 23, 2006}

\begin{abstract}
We explore the impact of new $SU(3) \times SU(2) \times U(1)$ invariant interactions characterized by a scale of order a TeV on Higgs boson properties. The Higgs production rate and branching ratios can be very different from their standard model values. We also discuss the possibility that these new interactions contribute to acceptable unification of the gauge couplings.
\end{abstract}

\maketitle

\section{Introduction}

The standard model for strong weak and electromagnetic interactions has provided an extremely successful description of experimental results.  The recent measurements of neutrino masses and mixings indicate new physics beyond that of the Glashow-Salam-Weinberg (GSW) theory. 

Many of the extensions of the standard model proposed in the literature have been motivated by the hierarchy puzzle.  The same naturalness arguments that make us uncomfortable about the smallness of the Higgs mass compared with the unification or Planck scales also apply to the cosmological constant. There is now experimental evidence for a cosmological constant of order $(10^{-3} {\rm eV})^4$, so Nature does not seem concerned about violations of naturalness.
 
If one does not use naturalness to motivate what new physics might occur at the weak scale, then a more  experimentally motivated approach is warranted. We shall  assume that the $SU(3) \times SU(2) \times U(1)$ GSW theory is valid at the electroweak symmetry breaking scale of order $ v \simeq 250$~GeV, and any new particles are at a mass scale heavier than the Higgs vacuum expectation value, so that their effects for Higgs physics can be parameterized in terms of $SU(3)\times SU(2) \times U(1)$ invariant higher dimension operators in the standard model. 

As we emphasize in this paper, it is possible that new physics associated with a mass scale well above the Higgs vacuum expectation value can have significant impact on the properties of the Higgs boson and still be consistent with the present experimental constraints on extensions of the standard model.\footnote{Models where the Higgs decays to new invisible light degrees of freedom have been studied for example in Refs.~\cite{Davoudiasl,Belotsky}. In our work there are no additional light degrees of freedom beyond those in the standard model.}
As a simple example, consider adding to the standard model the dimension six operator,
\begin{equation}
\label{example}
\delta {\cal L}=-{c_G g_3^2\over 2 \Lambda^2} H^{\dagger} H G^A_{\mu \nu} G^{A \mu \nu},
\end{equation}
where $G^{A \mu \nu}$ is the gluon field strength tensor and $H$ is the Higgs doublet. Expanding about the Higgs vacuum expectation value $v \simeq 250\,{\rm GeV}$,
\begin{eqnarray}
\delta {\cal L}&=& -{c_G g_3^2 v^2 \over 4 \Lambda^2} G^A_{\mu \nu}G^{A \mu \nu} -{c_G g_3^2 v h \over 2\Lambda^2} G^A_{\mu \nu} G^{A \mu \nu}\nn
&&-{c_G g_3^2 h^2 \over 4\Lambda^2} G^A_{\mu \nu} G^{A \mu \nu}.
\label{new}
\end{eqnarray}
The first term in Eq.~(\ref{new}) has the same form as the gluon kinetic term. It can be eliminated by rescaling the gluon field, and so can be absorbed into a shift of the gluon coupling constant.  As a result, the coefficient of the operator in Eq.~(\ref{example}) is only constrained by Higgs boson physics. The term linear in $h$ in Eq.~(\ref{new}) contributes to the production rate for Higgs bosons. For $m_h=120~{\rm GeV}$, this correction increases the production rate by about $20\%$ if $c_G$ is negative and $\Lambda/v \simeq 50{\sqrt c_G }$. Sizeable changes (i.e.\ big enough to be relevant for LHC Higgs searches) in the properties of the Higgs boson are possible even if the scale of new physics $\Lambda$ is greater than a $\rm TeV$. The reason for this is that in the standard model, the production cross section $\sigma^{\rm SM}(gg \rightarrow h)$ results from a one-loop matrix element $gg \rightarrow h $, and so has enhanced sensitivity to new physics.

The new physics effects we consider in this article are those which can significantly modify  the $gg \to h$ production rate and the $h \to gg$, $h \to \gamma \gamma$ and $h \to \gamma Z$ decay rates, all of which have one loop standard model amplitudes. 
In section~\ref{sec:ops}, we write down the dimension six operators that are important for the $h \to gg, \gamma\gamma,\gamma Z$ amplitudes. The experimental implications of these operators for Higgs production and decay is discussed in sections~\ref{sec:rates} and \ref{sec:exp}. The results given in these sections are general, and only depends on the existence of $SU(3)\times SU(2) \times U(1)$ invariant operators at some scale $\Lambda$ above the weak scale $v$. There are several hints for the existence of a unified theory at a high scale of order $10^{15}$~GeV. In section~\ref{sec:gut}, we examine the implications of unification for the size of the higher dimension operators. Conclusions are given in section~\ref{sec:conc}.

\section{Operators} 
\label{sec:ops}

We are interested in non-renormalizable operators that, when added to those in the minimal standard model, change the rates for the processes $gg \rightarrow h$, $h \rightarrow \gamma \gamma$ and $h \rightarrow Z \gamma$. The rates for these processes are measurable at the ${\rm LHC}$ and are sensitive to beyond the standard model physics since their standard model amplitudes start at one loop.\footnote{Strictly speaking, it is the Higgs production cross section times branching ratios for the above decays that are measurable. However, the total width for Higgs decay is approximately independent of the contributions of the higher dimension operators we are considering since the total width is dominated by the channels, $h \rightarrow b \bar b$, $h \rightarrow Z Z^*$ and $h \rightarrow W W^*$ which arise in the standard model from tree level matrix elements.} The relevant dimension six operators are:
\begin{eqnarray}
\label{lagrange}
&&\delta{\cal L}=-{c_G g_3^2\over 2 \Lambda^2} H^{\dagger} H G^A_{\mu \nu} G^{A \mu \nu}-{c_W g_2^2\over 2 \Lambda^2} H^{\dagger} H W^a_{\mu \nu} W^{a \mu \nu}\nn
&&-{c_B g_1^2\over 2 \Lambda^2} H^{\dagger} H B_{\mu \nu} B^{ \mu \nu}
- {c_{W\!B} g_1g_2\over 2 \Lambda^2} H^{\dagger}\tau^a H B_{\mu \nu} W^{a \mu \nu}\nn
&&-{\tilde c_G g_3^2\over 2 \Lambda^2} H^{\dagger} H \tilde G^A_{\mu \nu} G^{A \mu \nu}-{\tilde c_W g_2^2\over 2 \Lambda^2} H^{\dagger} H \tilde W^a_{\mu \nu} W^{a \mu \nu}\nn
&&-{\tilde c_B g_1^2\over 2 \Lambda^2} H^{\dagger} H \tilde B_{\mu \nu} B^{ \mu \nu}
- {\tilde c_{W\!B} g_1g_2\over 2 \Lambda^2} H^{\dagger}\tau^a H \tilde B_{\mu \nu} W^{a \mu \nu}
\end{eqnarray}
where $g_1$ and $g_2$ are the weak hypercharge and $SU(2)$ gauge couplings, $B^{\mu\nu}$ is the field strength tensor for the hypercharge gauge group and $W^{a \mu \nu}$ is the field strength tensor for the weak $SU(2)$ gauge group. In Eq.~(\ref{lagrange}) $\tilde G^A_{\mu \nu}=(1/2)\epsilon_{\mu \nu \lambda \sigma}  G^{A \lambda \sigma}$, etc.\ denote the duals of the field strength tensors.
The first four operators in Eq.~(\ref{lagrange}) conserve $CP$, and the last four violate $CP$. The $CP$ violating amplitudes do not interfere with the standard model amplitudes for Higgs decay. A complete list of dimension six operators can be found in Ref.~\cite{buchmuller}. The influence of these operators on some aspects of Higgs boson production was considered in Ref.~\cite{Rainwater,Han:2005pu,Logan:2004hj}.

Most of the terms in Eq.~(\ref{lagrange}) are not constrained by precision electroweak physics, which is mainly sensitive to corrections to the gauge boson propagators.
If we replace the Higgs field by its vacuum expectation value, the first three terms in Eq.~(\ref{lagrange}) just give a redefinition of the gauge couplings. The $c_{W\!B}$ term produces kinetic mixing of the $W_3$ and $B$ fields, and contributes to the variable $S$~\cite{Han},
\begin{equation}
c_{W\!B}=-{1 \over 8 \pi} {\Lambda^2 \over v^2}S.
\end{equation}
The $CP$ violating terms are discussed at the end of this section.

The experimental value of the $S$-parameter is $S=-0.13 \pm 0.10$ for $m_h=117$~GeV, and it decreases modestly as the Higgs mass increases. Suppose the central value $S=-0.13$ was due to the operator with coefficient $c_{W\!B}$. This implies that $\Lambda/v \simeq 14{\sqrt{c_{W\!B}}}$ and that $c_{W\!B}$ is positive. Consistency with precision electroweak physics is not a strong constraint. For example, if $c_G$ is the same magnitude, $\Lambda/v \simeq 14{\sqrt{c_{G}}}$,  then the production rate $\sigma^{\rm SM}(gg \rightarrow h)$ is almost zero if $c_G$ is positive, and $4$ times its standard model value if $c_G$ is negative! Since there can be very large enhancements of the Higgs production rate, one can place constraints on the $c$'s from existing Tevatron data.

Naive dimensional analysis~\cite{nda} suggests that the coefficients $c_j$ are of order $g_Y^2/(16{\pi^2})$, a value obtained by estimating the size of loop graphs. Here $g_Y$ is a coupling constant of the new heavy degrees of freedom in the loop to the Higgs doublet. This estimate holds even if the new physics is non-perturbative. However, there can be numerical enhancements associated with the number of degrees of freedom that contribute to loop graphs. For example, the degrees of freedom associated with the new physics could come in three generations just like the ordinary quarks and leptons. Consequently, for $g_Y \sim 1$, values for $|c_j|$ of order $0.1$ are reasonable. We will use values in the range $0.01-0.1$ when discussing the numerics. Coefficients of order $0.1$ play a role in the meeting of the couplings if there is unification, and can easily arise if the new physics degrees of freedom are in complete multiplets of a unification group --- see section~\ref{sec:gut}.

The typical values for $c$ that we use are smaller than those needed for observable effects in other gauge boson amplitudes that have been studied. Extensive studies have been made of the anomalous $WW\gamma$ couplings $\lambda_\gamma$ and $\Delta\kappa_\gamma$~\cite{Hagiwara}. The same naive dimensional estimates used above give
\begin{eqnarray}
\lambda_\gamma &\sim& g_2^2 \left( \frac{M_W}{\Lambda}\right)^2 c
 \sim \left(2 \times 10^{-3}\right) c,\nn
\Delta\kappa_\gamma &\sim& g_2^2 \left( \frac{v}{\Lambda}\right)^2 c 
\sim \left(2 \times 10^{-2}\right) c,
\end{eqnarray}
where we have used $\Lambda=1\,\text{TeV}$. Current experimental limits on these parameters are $-0.86 < \Delta\kappa_\gamma < 0.96$ and
$\abs{\lambda_\gamma} < 0.2$~\cite{D0}. Observable effects in anomalous gauge couplings would correspond to $c$'s of order 100, a factor of thousand larger than the $c$'s we are considering, which assumed $g_Y \sim 1$. Higgs production and decay rates provide a much more sensitive test for new physics than these anomalous gauge couplings.

The $CP$ violating terms in Eq.~(\ref{lagrange}) with the Higgs field replaced by its vacuum expectation value become topological terms such as $G \tilde G$, which are total derivatives,  and do not contribute in perturbation theory. They do contribute to the $\theta$ angle, and hence to the neutron electric dipole moment. The $CP$ violating terms in Eq.~(\ref{lagrange}) expanded to linear or quadratic order in the Higgs field generate $CP$ violating operators  such as $G \tilde G$ and $GG\tilde G$~\cite{weinberg} when heavy particles such as the Higgs and top quark are integrated out. The $G \tilde G$ operator also contributes to the $\theta$ angle and the neutron electric dipole moment. We assume that the entire $G \tilde G$ contribution to $\theta$ is removed by whatever mechanism solves the strong CP problem, otherwise it generates a neutron electric dipole moment which is experimentally ruled out by several orders of magnitude. Axion models, for example, eliminate the net low energy value of $\theta$, and so remove any contributions to $\theta$ from Eq.~(\ref{lagrange}). The dimension-six $GG\tilde G$ operator also contributes to the neutron electric dipole moment. Following the analysis of Ref.~\cite{weinberg}, ignoring any anomalous dimensions, and using naive dimensional analysis, we estimate that the induced neutron electric dipole moment from the operator in Eq.~(\ref{lagrange}) involving $G \tilde G$ is $d \sim 10^{-25} e\text{-cm}\times \tilde c_G \times (1\,\text{TeV}/\Lambda)^2$. This is consistent with the current experimental limits for the range of parameters considered here. Similar analyses hold for the other $CP$ violating operators in Eq.~(\ref{lagrange}).  

\section{Higgs Production and Decay Rates}
\label{sec:rates}

The operators in Eq.~(\ref{lagrange}) lead to additional vertices of the form $h G^A_{\mu \nu} G^{A \mu \nu}$, $h W^a_{\mu \nu} W^{a \mu \nu}$, etc.\  when $H$ is expanded in powers of the Higgs field. These additional local vertices modify the Higgs decay amplitude, which we compute in this section.

The decay $h \rightarrow gg$ is not observable at the LHC. For a  Higgs in the mass range we are considering, its dominant production mechanism at the LHC is via $gg \to H$. In the standard model, the dominant $gg \to h$ (or $h \to gg$) amplitude is due to a virtual top quark loop. The $gg \to h$ amplitude, even for a 120~GeV Higgs, is well-approximated by the amplitude in the $m_t \to \infty$ limit (for a review see Ref.~\cite{reina}). In the $m_t \to \infty$ limit, the standard model $gg \to h$ amplitude is given by a local $h G^A_{\mu \nu} G^{A \mu \nu}$ operator, so the QCD radiative corrections from scales below $m_t$ to the standard model amplitude and the new physics correction are identical, since both amplitudes have the same operator form. 
For this reason, we will give the ratio of the modified production and decay rates to their standard model value. At present the $gg \to h$ production mechanism has been calculated to NNLO order and a soft gluon resummation has been done \cite{nlo,nnlo,resum}. The theoretical uncertainty from higher order QCD corrections is estimated to be around $10\%$. There are also additional sources of uncertainly in the theoretical prediction for the cross section, for example, from the uncertainty in the parton distributions and the top quark mass.

\begin{widetext}

\subsection{$gg \to h$ or $h \to gg$}

The ratio is
\begin{eqnarray}
\label{glue}
{ \sigma(gg \to h) \over \sigma^{\text{SM}}(gg \to h)} \simeq { \Gamma(h\to gg) \over \Gamma^{\text{SM}}(h \to gg)}
&\simeq& \abs{1- {8 \pi^2 v^2 c_G \over \Lambda^2 I^g}}^2+\abs{ {8 \pi^2 v^2 \tilde c_G \over \Lambda^2 I^g}}^2
\end{eqnarray}
where the standard model amplitude is dominated by integrating out the top quark 
\begin{eqnarray}
I^g &=& I_f(m_h^2/(4 m_t^2),0)\left(1+ \frac{11}{4}{\alpha_s \over \pi }\right),
\label{igamp}
\end{eqnarray}
with
\begin{eqnarray}
I_f(a,b) &=&\int_0^1 {\rm d}x \int_0^{1-x} \hspace{-0.3cm}{\rm d}y\ {1-4x y \over 1 -4(a-b) xy- 4 b y(1-y)-i0^+}.
\label{if}
\end{eqnarray}
using the notation of Ref.~\cite{BH} for the parameter integral.\footnote{Note that $f_2 \to -f_2 $ in the explicit evaluation of the parameter integral given in Eq.~(4) of Ref.~\cite{BH}} In Eq.~(\ref{glue}) we have included the standard model two loop matching~\cite{match,match1} onto the operator $h G^A_{\mu \nu} G^{A \mu \nu}$ in the large $m_t$ limit,  multiplied by the exact one loop expression for the amplitude $I_f$. This expression gives the full $m_h/m_t$ dependence of the leading term, and gives the $\alpha_s$ correction for $m_t \to \infty$. The $\alpha_s$ correction is numerically about 10\%. The $\alpha_s^2$ and $\alpha_s^3$ corrections for the $hgg$, $h\gamma\gamma$ and $h \gamma Z$ amplitudes are also known~\cite{match1}. Their contributions are smaller, and not included here.
Radiative corrections from initial state gluon radiation, etc.\ cancel out in the ratio Eq.~(\ref{glue}) in the $m_t \to  \infty$ limit.
We have also neglected the running of the coefficient $c_G$ between the scale of new physics $\Lambda$ and the top quark mass.  This correction can be computed, and is about 2\%.

Two gluons can also produce a Higgs via associated production off heavy quarks \cite{heavyquarkt,heavyquarkb}, i.e., $gg \rightarrow Q \bar Q h$. Eq.~(\ref{glue}) does not include this process. The amplitude for associated Higgs production off of heavy quarks starts at tree level and so we expect it to be largely unaffected by new physics effects at the TeV scale.

\subsection{$h \to \gamma \gamma$}
The ratio is
\begin{eqnarray}
\label{phot}
{ \Gamma(h \to \gamma \gamma) \over \Gamma^{\text{SM}}(h \to \gamma \gamma)}
&\simeq& \abs{1- {4 \pi^2 v^2 c_{\gamma \gamma}\over \Lambda^2 I^\gamma}}^2+
\abs{ {4 \pi^2 v^2 \tilde c_{\gamma \gamma}\over \Lambda^2 I^\gamma}}^2 ,
\end{eqnarray}
where $c_{\gamma \gamma}=c_W+c_B -c_{W\!B}$, $\tilde c_{\gamma \gamma}=\tilde c_W+\tilde c_B -\tilde c_{W\!B}$. The standard model amplitude is given by
\begin{eqnarray}
I^\gamma &=& I_W^\gamma(m_h^2/(4m_W^2),0) +  N_c Q_t^2 I_f(m_h^2/(4 m_t^2),0)\left(1-{ \alpha_s \over \pi }\right),
\label{igamma}
\end{eqnarray}
where the fermion contribution is dominated by the top quark loop.
The Feynman parameter integral $I_W^\gamma$ is defined as in Ref.~\cite{BH},
\begin{eqnarray}
I^\gamma_W(a,b) &=& \int_0^1 {\rm d}x \int_0^{1-x} \hspace{-0.3cm}{\rm d}y\ {-4+6 xy+4 a x y \over 1 -4(a-b) xy- 4 b y(1-y)-i0^+}.
\label{igw}
\end{eqnarray}
In Eq.~(\ref{igamma}) $N_c=3$ is the number of colors and $Q_t=2/3$ is the top quark charge. An analytic expression for the Feynman parameter integration $I^\gamma_W(a,b)$ is known~\cite{BH}.

\subsection{$h \to \gamma Z$}

The ratio is
\begin{eqnarray}
\label{z}
&&{ \Gamma(h \to \gamma Z) \over \Gamma^{\text{SM}}(h \to \gamma Z)}\simeq\abs{1- {4 \pi^2 v^2 c_{\gamma Z}\over \Lambda^2 I^Z}}^2+
\abs{{4 \pi^2 v^2 \tilde c_{\gamma Z} \over \Lambda^2 I^Z}}^2 ,
\end{eqnarray}
where $c_{\gamma Z}=c_W \cot \theta_W -c_B \tan \theta_W -c_{W\!B} \cot 2 \theta_W$,  $\tilde c_{\gamma Z}=\tilde c_W \cot \theta_W -\tilde c_B \tan \theta_W -\tilde c_{W\!B} \cot 2 \theta_W $, and the standard model amplitude is
\begin{eqnarray}
I^Z &=& I_W^Z(m_h^2/(4m_W^2),m_Z^2/(4m_W^2)) +  N_c Q_t  g_t I_f(m_h^2/(4 m_t^2),m_Z^2/(4m_t^2))\left(1-{ \alpha_s \over \pi }\right),
\label{izamp}
\end{eqnarray}
with $g_f=( T_{3f} -2 \sin^2 \theta_W Q_f)/\sin 2\theta_W$, $T_{3t}=1/2$ the weak isospin of the left handed top quark, and $I_W^Z$ defined as in Ref.~\cite{BH},
\begin{eqnarray}
I^Z_W(a,b) &=& {1\over \tan \theta_W} \int_0^1 {\rm d}x \int_0^{1-x} \hspace{-0.3cm}{\rm d}y\ {\left[5 - \tan^2 \theta_W+2 a\left(1 - \tan^2 \theta_W \right)\right]xy-\left( 3 - \tan^2 \theta_W\right)\over 1 -4(a-b) xy- 4 b y(1-y)-i0^+}.
\label{izw}
\end{eqnarray}
We have only included the dominant fermionic contribution to $I^Z$ which comes from the top quark. Note that $N_c Q_t  g_t=(3-8 \sin^2\theta_W)/(3 \sin 2\theta_W) $. An analytic expression for the Feynman parameter integration $I^Z_W(a,b)$ is known~\cite{BH}.

\end{widetext}

\begin{figure}
\begin{center}
\includegraphics[bb=18 320 592 718,width=8cm]{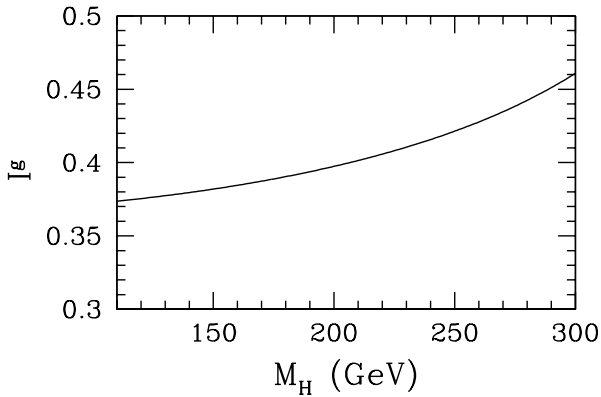}
\end{center}
\vspace{-0.5cm}
\caption{The standard model $h \to gg$ amplitude $I^g$ given by Eq.~(\ref{igamp}) plotted as a function of the Higgs mass.\label{fig:gluon}}
\end{figure}

\begin{figure}
\begin{center}
\includegraphics[width=8cm]{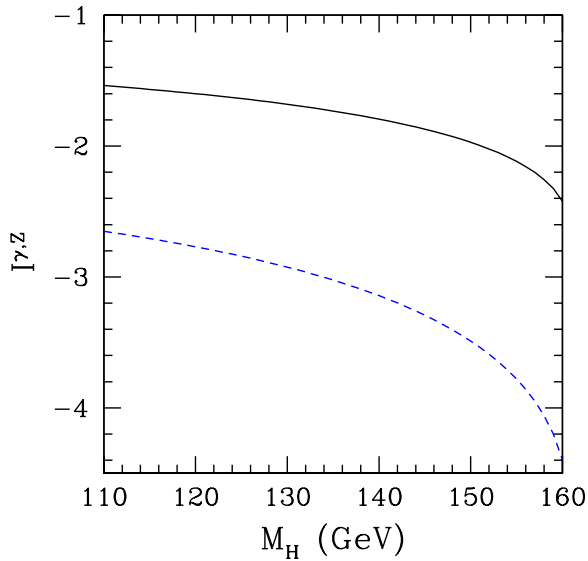}
\end{center}
\vspace{-0.75cm}
\caption{The standard model $h \to \gamma\gamma$ amplitude $I^\gamma$ given by Eq.~(\ref{igamma}) (solid curve) and the $h \to \gamma Z$ amplitude $I^Z$ given by Eq.~(\ref{izamp}) (dashed curve) plotted as a function of the Higgs mass. \label{fig:photon}}
\end{figure}

\section{Numerics}
\label{sec:exp}

We will use the PDG~\cite{pdg} central values $M_W=80.425$~GeV, $M_Z=91.1876$~GeV, $m_t=174.3$~GeV, $\sin^2\theta_W = 0.23120$, and $\alpha^{-1}(M_Z)=127.918$ for our numerical work. The standard model amplitudes, $I^g$, $I^{\gamma}$, $I^{Z}$ are plotted in Figs.~(\ref{fig:gluon}) and~(\ref{fig:photon}), as a function of the Higgs mass. For the radiative correction, we have used $\alpha_s \to \alpha_s(m_t)=0.108$.
Since their dependence on the Higgs mass is weak, simple approximation formul\ae\ are possible for the quantities $1/I^g$, $1/I^{\gamma}$ and $1/I^Z$. In the region $110 \le m_h \le 300$~GeV,
\begin{eqnarray}
{1 \over I^g} = 2.76 - 6.37 \times 10^{-2} \left( {m_h \over 100\ \text{GeV}}\right)^2,
\label{approxgluon}
\end{eqnarray}
with a maximum error less than 0.02, and in the region $110 \le m_h \le 160$~GeV,
\begin{eqnarray}
{1 \over I^\gamma} &=& -0.85 + 0.16  \left( {m_h \over 100\ \text{GeV}}\right)^2, \nn[5pt]
{1 \over I^Z} &=& -0.51 + 0.10  \left( {m_h \over 100\ \text{GeV}}\right)^2 ,
\end{eqnarray}
with errors less than $0.04$.

\begin{figure}
\begin{center}
\includegraphics[width=8cm]{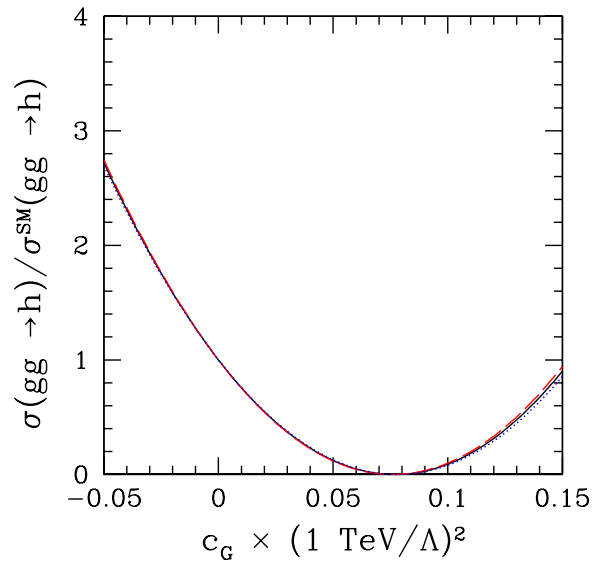}
\end{center}
\vspace{-0.75cm}
\caption{The ratio of the $gg \to h$ cross-section to its standard model value as a function of $c_G$ for $m_h=120$~GeV (dashed red), $m_h=140$~GeV (solid black) and $m_h=160$~GeV (dotted blue). The variation with Higgs mass is very small. \label{fig:grates}}
\end{figure}

$I^g$ is the smallest in magnitude, so if all the coefficients $c_j$ were the same size, the new physics would be most important for the $gg \to h$ cross section. In Fig.~(\ref{fig:grates}) we plot the $gg \to h$ ratio in Eq.~(\ref{glue}) as a function of $c_G \times (1\, {\text{TeV}}/\Lambda)^2$ with $\tilde c_G=0$ for $m_h=120,140,160$~GeV. 
The $gg \to h$ rate depends very weakly on the Higgs mass in this range. The new physics contributions make a dramatic difference to the production rate.
For $\Lambda=1$~TeV and $c_G=0.01$ the production cross section is $76\%$ of its standard model value and for $c_G=-0.05$ it is 3 times its standard model value. 

\begin{figure}
\begin{center}
\includegraphics[width=8cm]{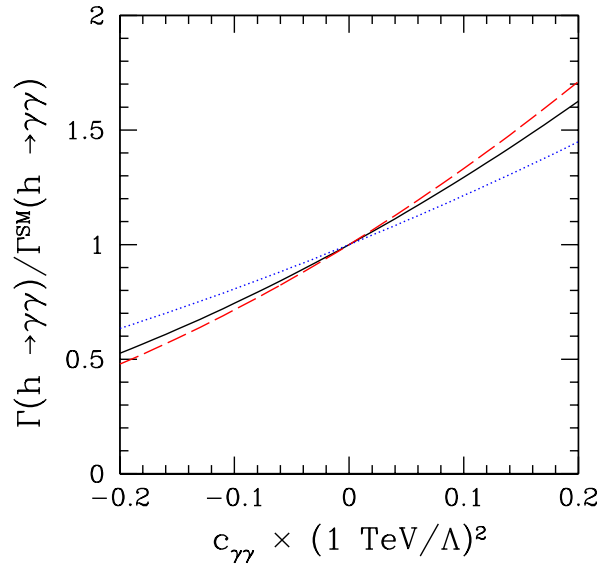}
\end{center}
\vspace{-0.75cm}
\caption{The ratio of the $h \to \gamma\gamma$ decay rate to its standard model value as a function of $c_{\gamma\gamma}$ for $m_h=120$~GeV (dashed red), $m_h=140$~GeV (solid black) and $m_h=160$~GeV (dotted blue). \label{fig:prates}}
\end{figure}

\begin{figure}
\begin{center}
\includegraphics[width=8cm]{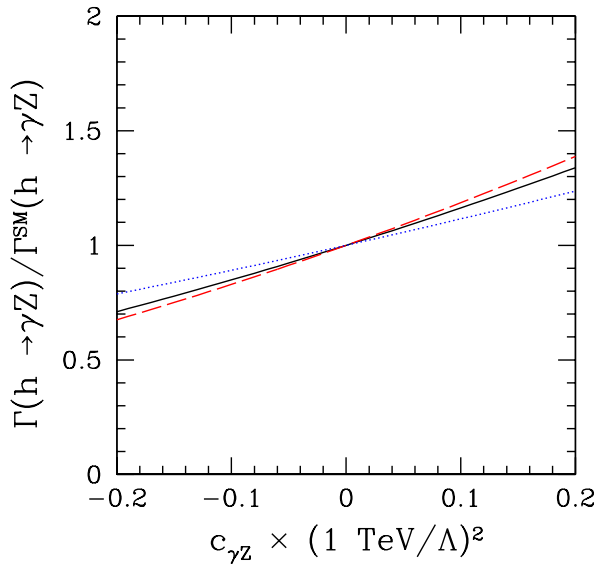}
\end{center}
\vspace{-0.75cm}
\caption{The ratio of the $h \to \gamma Z$ decay rate to its standard model value as a function of $c_{\gamma Z}$ for $m_h=120$~GeV (dashed red), $m_h=140$~GeV (solid black) and $m_h=160$~GeV (dotted blue). \label{fig:zrates}}
\end{figure}

 In Figs.~(\ref{fig:prates}) and (\ref{fig:zrates}) we plot the $h \to \gamma\gamma$ and $h \to \gamma Z$ ratios in Eqs.~(\ref{phot}) and (\ref{z}) as a functions of $c_{\gamma \gamma} \times (1\, {\text{TeV}}/\Lambda)^2$ and $c_{\gamma Z} \times (1\, {\text{TeV}}/\Lambda)^2$ respectively, when $\tilde c_{\gamma \gamma}=\tilde c_{\gamma Z}=0$ and $m_h=120,140,160$~GeV. 
There is a much larger dependence on the Higgs mass than for $gg \to h$.
Consider the choice of parameters $m_h=140$~GeV, $\Lambda=1$~TeV and $c_W=c_B=0.01,~c_{W\!B}=0$, which are small values for the $c$'s,  and correspond to $c_{\gamma \gamma}=0.02$ and $c_{\gamma Z}\simeq 0.013$.
The Higgs decay rate to two photons is increased by $6\% $ over its standard model value and the Higgs decay rate to a photon and a $Z$ boson is increased by roughly $2\%$ over its standard model value. For larger values of the coefficients $c_{\gamma \gamma}$ and $c_{\gamma Z}$, the rates can be changed substantially from their standard model values. If $c_{\gamma \gamma}=c_{\gamma Z}=0.1$, the $h \to \gamma\gamma$ and $h \to \gamma Z$ rates are 1.30 and 1.16 times their standard model values.

The event rate for $\gamma\gamma$ and $\gamma Z$ final states, which will be used to search for Higgs bosons with $m_h < 2 M_W$, depend on the product $\sigma(gg \to h) \Gamma(h \to \gamma \gamma)$ and $\sigma(gg \to h) \Gamma(h \to Z \gamma)$.

For $m_h >2 M_W$, the dominant signal is via the $h \to WW$ decay mode, which has a tree-level amplitude in the standard model. The new physics contributions are much smaller than the tree-level amplitude, so $\Gamma(h \to WW)$ is virtually unaffected by new physics contributions. The $WW$ event rate is still sensitive to new physics via the production cross-section $\sigma(gg \to h)$.

\section{Unification as a Guide to the Size of Coefficients}
\label{sec:gut}

The smallness of neutrino masses suggests that, in the GSW model, they are generated via non-renormalizable dimension five operators. Such dimension five operators arise in unified theories via the seesaw mechanism~\cite{seesaw}, and the size of observed neutrino masses is consistent with this idea if the unification scale is of order $10^{15}$~GeV.  The approximate unification of the coupling constants in the GSW model when evolved to high energies also suggests the existence of a unified theory at some high scale of order $10^{15-19}$~GeV, e.g. $SO(10)$ with fermions in the $\bf{16}$ or $SU(5)$ with fermions in the reducible representation $\bf{1+ \bar 5 + 10}$.

In unified theories with a high unification scale, triviality of $\lambda \phi^4$ theory implies that the Higgs must be relatively light~\cite{triviality}. The  Higgs mass upper bound for a unification scale of $10^{15}$~GeV  is about 170~GeV~\cite{upperbound}.
Another indication that the Higgs boson is light is from a fit to all the precision electroweak data~\cite{pdg}. For these reasons, the $m_h < 2 M_W$ region, which we concentrate on this paper, is particularly interesting to study.

New physics gives rise to shifts in the gauge couplings, for example from the first term in Eq.~(\ref{new}). Since the low-energy values of the couplings are fixed by experiment, this implies a shift in the high-energy values of the coupling. This, in turn, affects the unification of the coupling constants. In this section we explore what sizes of the coefficients, such as $c_G$, have significant impact on the condition that coupling constant unification holds. If the degrees of freedom associated with new physics come in complete $SU(5)$ multiplets,\footnote{Note that $SO(10)$ representations form complete $SU(5)$ multiplets.} then they will not affect the running at the leading logarithmic level, but will through threshold effects such as $c_G$. We will assume complete $SU(5)$ multiplets for the purposes of this analysis, for simplicity. It is straightforward to eliminate this assumption, but the results then depend on the particle content assumed for the new physics.

The new terms that are relevant for coupling constant unification are dimension four operators
\begin{equation}
\label{renorm}
\delta{\cal L}=-{\epsilon_G g_3^2 \over 2}G^A_{\mu \nu} G^{A \mu \nu}-{\epsilon_W g_2^2 \over 2}W^a_{\mu \nu} W^{a \mu \nu}-{\epsilon_B g_1^2 \over 2}B_{\mu \nu} B^{ \mu \nu}.
\end{equation}
We expect the $c$'s in Eq.~(\ref{lagrange}) to be of order $g_Y^2$ times the corresponding $\epsilon$'s in Eq.~(\ref{renorm}). Rescaling the gauge fields to remove these terms changes the gauge couplings from their values in standard unification $(SU)$ to their measured values, $g_1, g_2, g_3$. To first order in $\epsilon_{G,W,B}$:
\begin{eqnarray}
g_1&=&g_1^{(SU)}\left[1-\left({4 \pi \alpha_e \over {\rm cos}^2\theta_W}\right)\epsilon_B\right],\nn[5pt]
g_2&=&g_2^{(SU)}\left[1-\left({4 \pi \alpha_e \over {\rm sin}^2\theta_W}\right)\epsilon_W\right],\nn[5pt]
g_3&=&g_3^{(SU)}\left[1- 4 \pi \alpha_s\epsilon_G\right],
\label{su1}
\end{eqnarray}
or equivalently,
\begin{eqnarray}
\alpha_e &=& \alpha_e^{(SU)}\left[1-8 \pi \alpha_e \left(\epsilon_B+\epsilon_W\right)\right],\nn[5pt]
\sin^2\theta_W &=&\sin^2\theta_W^{(SU)}+8 \pi \alpha_e \left( \cos^2 \theta_W \epsilon_W - \sin^2 \theta_W \epsilon_B \right) \nn[5pt]
\alpha_s &=& \alpha_s^{(SU)}\left[1- 8 \pi \alpha_s \epsilon_G\right].
\label{su2}
\end{eqnarray}
Here a superscript $(SU)$ denotes the value in standard unification without the modifications that Eq.~(\ref{renorm}) cause to the gauge couplings.

At the leading logarithmic level standard unification predicts that,
\begin{eqnarray}
\sin^2\theta_W^{(SU)}\left(M_Z\right)={3(b_3-b_2) \over \Delta_b }+{(5b_2-3b_1) \over \Delta_b}\left[{ \alpha_e^{(SU)}(M_Z) \over \alpha_s^{(SU)}(M_Z)}\right]\nn
\label{1}
\end{eqnarray}
and
\begin{eqnarray}
{\rm ln}\left({M_{G} \over M_Z}\right)={6\pi \over \Delta_b}\left[{1 \over \alpha_e^{(SU)}(M_Z)}-{8 \over 3}{1 \over \alpha_s^{(SU)}(M_Z)}\right]
\label{2}
\end{eqnarray}
where $M_G$ is the unification scale, $\Delta_b = 8b_3-3b_1-3b_2$,
\begin{eqnarray}
b_1 &=& 0-\frac {20} 9 n_g - \frac 1 {6},\nn
b_2 &=& \frac {22} 3 - \frac 4 3 n_g - \frac 1 6, \nn
b_3 &=& 11 - \frac  4 3 n_g+0,
\label{bcoeff}
\end{eqnarray}
are the coefficients of the one-loop $\beta$-functions,
\begin{eqnarray}
\mu {{\rm d} g_i \over {\rm d}\mu} &=& - {b_i \over 16 \pi^2} g_i^3 + \ldots .
\label{beta}
\end{eqnarray}
The three terms in Eq.~(\ref{bcoeff}) are the gauge boson, fermion and Higgs doublet contributions, respectively, and $n_g$ is the number of generations. Since we are neglecting matching corrections at the GUT scale we imposed the conditions,
\begin{equation}
g_3^{(SU)}(M_G)=g_2^{(SU)}(M_G)={\sqrt {5 \over 3}} g_1^{(SU)}(M_G).
\end{equation}

Using the relations between the measured gauge couplings and their values in standard unification, Eqs.~(\ref{su1},\ref{su2}) and neglecting any running between the scale $\Lambda$ and $M_Z$, the relations Eqs.~(\ref{1},\ref{2}) imply that
\begin{eqnarray}
&&{\rm ln}\left({M_{GUT} \over M_Z}\right)={6\pi \over \Delta_b}\left[{1 \over \alpha_e(M_Z)}-{8 \over 3}{1 \over \alpha_s(M_Z)}\right]\nn
&&+ {48 \pi^2 \over \Delta_b} \left[ \frac 8 3 \epsilon_G - \epsilon_W - \epsilon_B \right]\nn
&\simeq& 29.67+18.9 \epsilon_G - 7.1 \left(\epsilon_W +\epsilon_B\right)
\label{mg}
\end{eqnarray}
\begin{eqnarray}
&&\sin^2\theta_W={3(b_3-b_2) \over \Delta_b}+{(5b_2-3b_1)\over \Delta_b}\left[{ \alpha_e(M_Z) \over \alpha_s(M_Z)}\right] 
 \nn
 &&+ 8 \pi \alpha_e \Biggl\{{ (5b_2-3b_1)\over \Delta_b}
 \left( \left[{ \alpha_e(M_Z) \over \alpha_s(M_Z)}\right]\left[\epsilon_B + \epsilon_W\right] -\epsilon_G\right)\nn
 &&+\left[\cos^2 \theta_W \epsilon_W - \sin^2 \theta_W \epsilon_B\right]\Biggr\}\nn
 &\simeq& 0.2074-0.107 \epsilon_G + 0.158 \epsilon_W -  0.038 \epsilon_B.
\end{eqnarray}
If we require that $\sin^2 \theta_W (M_Z) = 0.23120$, we get the constraint
\begin{eqnarray}
1 &=& 6.63 \epsilon_W - 1.61 \epsilon_B - 4.47 \epsilon_G.
\label{c1}
\end{eqnarray}

One of the problems with minimal $SU(5)$ unification is the proton lifetime. If the unification scale is raised by a factor $f$ over the minimal $SU(5)$ unification scale, with $f \agt 10$, then the proton lifetime is consistent with the experimental limits. Using Eq.~(\ref{mg}), this gives the constraint
\begin{eqnarray}
\ln f & = & 18.9 \epsilon_G - 7.1 \left(\epsilon_W +\epsilon_B\right)
\label{c2}
\end{eqnarray}
with $\ln f \agt 2.3$.

The constraints in Eqs.~(\ref{c1}) and (\ref{c2}) can be used to get an estimate of the sizes of $\epsilon_{G,W,B}$ that can contribute significantly to improving the meeting of the couplings. If there are no large cancellations between the various contributions, then $\epsilon_{G,W,B}$ and hence (for $g_Y \sim 1$) $c_{G,W,B}$ with magnitudes of order $0.1$ will play a role in unification of the gauge couplings. It has been noted previously that non-supersymmetric unification of the couplings can arise from GUT scale threshold corrections~\cite{gut}. GUT scale threshold corrections are given by terms such as Eq.~(\ref{renorm}) generated at the unification scale. Since Eq.~(\ref{renorm}) contains dimension four operators, there are no powers of $\Lambda$ in the denominator, and hence no information on the scale at which they are generated. For this reason, one can only use Eqs.~(\ref{c1}) and (\ref{c2}) as an estimate for $\epsilon_j$, rather than as a constraint.

For $g_Y \sim 1$, naive dimensional analysis~\cite{nda} predicts that $\epsilon_j \sim c_j \sim 1/(16 \pi^2)$. This estimate comes from examining the typical size of loop graphs. If there are many degrees of freedom contributing in loops, then the coefficients can be larger. A simple example that generates the operators in Eq.~(\ref{lagrange}) is to add a fermion multiplet in the ${\bf 16+\overline{16}}$ of $SO(10)$, with a mass $\Lambda$. The ${\bf 16 + \overline{16}}$ can have both a $SO(10)$ invariant mass term, and also couple to the ${\bf 10}$ of $SO(10)$ which contains the Higgs, so it generates Higgs operators such as those in Eq.~(\ref{lagrange}) when the fermions are integrated out. The index of the ${\bf 16 + \overline{16}}$ is 8 times that of a fermion in the fundamental representation, so it produces coefficients $\epsilon_j \sim c_j \sim 8/(16 \pi^2) \sim 0.1$.

\section{Conclusions}
\label{sec:conc}

Higgs bosons at the LHC are produced by gluon fusion, $gg \to h$.  The favored detection channel for $m_h < 2 M_W$ is via the $h \to \gamma \gamma$ rate. The $h \to \gamma Z$ decays can also be used to search for a Higgs lighter than $2M_W$.
 All three processes have amplitudes which start at one loop in the standard model, and hence are particularly sensitive to new physics. We used an operator analysis to examine the possible impact of such new physics on these processes. Most of the relevant operators are unconstrained by precision electroweak physics. The production rate $\sigma(g g \to h)$ and the decay rates $\Gamma (h \rightarrow g g,\gamma\gamma,\gamma Z)$ are very sensitive to beyond the standard model physics and dramatic changes in the production cross section are possible with the new physics at a scale around a ${\rm TeV}$ and coefficients in the Lagrangian density consistent with naive dimensional analysis. Tree-level standard model processes are much less sensitive to new physics contributions. By comparing the rates for one-loop and tree-level processes, one can determine whether there is a new physics contribution to Higgs interaction amplitudes. This allows one to rule out other possible explanations for anomalous rates, such as modifications to the parton distribution functions.
 
 The $\gamma \gamma$ and $\gamma Z$ decay channels will mainly be useful for a Higgs with $m_h < 2 M_W$.  The $gg \to H$ production channel dominates at the LHC, and so can be used to search for new physics, even if the Higgs is heavier than $2M_W$. The $gg \to H$ rate is enhanced if $c_G$ is negative and suppressed if $c_G$ is positive; if the suppression is large enough, other production channels can dominate.

\begin{acknowledgments}
  This work was supported in part by DOE grants   DE-FG03-97ER40546 and DE-FG03-92ER40701.
\end{acknowledgments}


\end{document}